\documentclass[aps,prb,twocolumn,floatfix,notitlepage, superscriptaddress,longbibliography]{revtex4-2}
\usepackage{amssymb,amsmath,amsfonts} 
\usepackage{newtxtext,newtxmath}
\usepackage[utf8]{inputenc}
\usepackage{amsfonts,amsmath,amssymb,graphicx,epstopdf,verbatim}
\usepackage{rotating}
\usepackage{epstopdf}
\usepackage{xcolor}
\usepackage{natbib}
\usepackage[colorlinks]{hyperref}
\definecolor{darkred}{rgb}{0.5,0,0}
\definecolor{darkgreen}{rgb}{0,0.5,0}
\definecolor{darkblue}{rgb}{0,0,0.5}
\hypersetup{colorlinks,
linkcolor=darkred,
filecolor=darkgreen,
urlcolor=darkblue,
citecolor=darkgreen}
\usepackage{braket}
\usepackage{overpic}
\usepackage{orcidlink}

\begin{document}

\title{Quantum many-body scars leading to time-translation symmetry breaking in kicked interacting spin models}

\author{\'Angel L. Corps~\orcidlink{0000-0002-0666-6642}}
    \email[]{corps.angel.l@gmail.com}
    \affiliation{Institute of Particle and Nuclear Physics, Faculty of Mathematics and Physics, Charles University, V Hole\v{s}ovi\v{c}k\'{a}ch 2, 180 00 Prague, Czech Republic}
    \affiliation{Grupo Interdisciplinar de Sistemas Complejos (GISC), Universidad Complutense de Madrid, Av. Complutense s/n, E-28040 Madrid, Spain}

\author{Armando Rela\~no~\orcidlink{0000-0002-1670-1544}}
    \email[]{armando.relano@fis.ucm.es}
    \affiliation{Grupo Interdisciplinar de Sistemas Complejos (GISC), Universidad Complutense de Madrid, Av. Complutense s/n, E-28040 Madrid, Spain}
    \affiliation{Departamento de Estructura de la Materia, Física Térmica y Electrónica, Universidad Complutense de Madrid, Av. Complutense s/n, E-28040 Madrid, Spain}

\author{Angelo Russomanno~\orcidlink{0009-0000-1923-370X}}
    \email[]{angelo.russomanno@unina.it}
    \affiliation{Dipartimento di Fisica ``E. Pancini'', Università di Napoli Federico II, I-80126 Napoli, Italy}


\begin{abstract}
We study an Ising model with long-range interactions undergoing a time-periodic kicking. For different initial states we observe persistent period doubling. When there is period  doubling we find that the initial state has relevant overlap with Floquet states showing time-translation symmetry breaking, organized in doublets displaying $\pi$-spectral pairing (as highlighted by the $\pi$-spectral gap) and long-range order (as shown by the eigenvalues of the magnetization in the doublet). We observe period doubling for initial states with domain walls and tilted spins, and for the latter ones a finite-size scaling of the relevant $\pi$-shifted gap and magnetization eigenvalues suggests period-doubling oscillations persisting for larger system sizes and lasting a time exponential in the system size. We find that just a minority of Floquet states displays time-translation symmetry breaking while the rest is thermal, a weak-ergodicity breaking situation typical of systems with quantum scars. Although the time-translation symmetry breaking eigenstates are the minority, their number is exponential in the system size and this motivates the period doubling observed for many different initial states.
\end{abstract}

\maketitle

\section{Introduction}
A flourishing research field in recent years has been the one of Floquet time crystals, where a periodically driven many-body quantum system provides in the thermodynamic limit a persistent response with a period multiple of the one of the driving (see the reviews~\cite{Yao_2018,annurev_Else}). Following the first proposals in Refs.~\cite{PhysRevLett.117.090402,PhysRevLett.116.250401}, experimental demonstrations~\cite{articlei_nat,Zhang_2017,Choi_2017} and realizations in the context of digital quantum computers~\cite{PRXQuantum.2.030346,2022Natur.601..531M,6b06a3bf9f9847ce8a9081b79ef82b07,sciadv.abm7652,Cae3851,ma2025quditnativeframeworkdiscretetime,shinjo2026noisestabilizeddiscretetimecrystals} have been presented. A key aspect is avoiding infinite-temperature thermalization in order to get persistent time-translation-symmetry breaking oscillations. This is usually achieved by adding disorder that hinders thermalization through many-body localization~\cite{khemani2019briefhistorytimecrystals}. 

In this framework, the properties of the Floquet eigenstates and quasienergies~\cite{PhysRevA.7.2203,PhysRev.138.B979} are crucial. They must be organized in doublets separated by a gap $\pi/\tau$, where $\tau$ is the period, to provide synchronized Rabi oscillations with a period twice that of the driving~\cite{PhysRevB.94.085112}. Moreover, they must be cat-state superpositions of dressed classical spin configurations, so that these Rabi oscillations occur between states with a global magnetization that changes sign at every period. This cat-state superposition feature, similar to the two degenerate ground states that break the $\mathbb{Z}_2$ symmetry in the Ising chain in transverse field, can be highlighted through the presence of long-range correlations~\cite{PhysRevLett.117.090402}. These two properties -- $\pi$-spectral pairing and long-range correlations -- are both crucial for having a time crystal. These are the ``time-translation symmetry breaking'' properties of the Floquet states, and occur all over the Floquet spectrum for proposals based on disorder and many-body localization.

In this context, proposals based on long range interactions~\cite{PhysRevX.12.031037,PhysRevA.99.033618,Pizzi_2021,PhysRevB.109.174310,PhysRevB.111.174311} are especially interesting. If the interactions are actually infinite range~\cite{Pizzi_2021,PhysRevB.95.214307}, ergodicity can be fully broken, and the Hilbert space is fragmented so that the whole Floquet spectrum is nonthermal and all eigenstates show time-translation symmetry breaking. The case of interactions decaying as a power-law in absence of disorder is even more interesting. In this case, without a periodic driving, the Hamiltonian displays weak ergodicity breaking~\cite{Lerose_2025,PhysRevB.104.094309,kv2w-mk4t}, in the sense that only a fraction of the eigenstates are nonthermal and one can see an absence of thermalization when the system is prepared in an initial state with large overlap with one of these states. It is remarkable that classical spin configurations are among these initial states leading to a nonthermalizing dynamics. These nonthermal eigenstates are called quantum scars and appear in many contexts. Initially discovered in the Rydberg atom simulator~\cite{Bernien_2017}, scarring phenomena have been explored in a broad class of systems, including the Rydberg-blockaded PXP models~\cite{2018NatPh..14..745T,Turner_2018,PhysRevB.99.161101,PhysRevLett.122.173401,Iadecola_2019,Choi_2019,Ho_2019,PhysRevB.101.245107,Surace_2021,paviglianiti2025}, the AKLT model~\cite{PhysRevLett.59.799,Shiraishi_2019,PhysRevB.98.235156}, Josephson-junction arrays~\cite{PhysRevB.106.035123}, spin-1 models~\cite{halder2025thermalizationexactquantummanybody,PhysRevB.108.104411,You_2022,Schecter_2019,4tv9-q7g7,PhysRevB.107.235121}, frustrated hardcore bosons~\cite{ding2026exactquantumscarsfrustrated}, multispecies bosonic Josephson junctions~\cite{dimenna2026}, qubit models with Kramers-Wannier duality~\cite{Fontana_2026}, and various fermionic models~\cite{PhysRevResearch.3.043156,PhysRevB.102.075132,PhysRevB.102.085140,PhysRevLett.125.230602,PhysRevResearch.5.043208,PhysRevResearch.6.043259,PRXQuantum.4.040348}.

An interesting question therefore is about the presence of time translation symmetry breaking in systems with quantum scars. One can imagine a driven system where most of the eigenstates are thermal -- in the sense that are locally equivalent to an infinite-temperature thermal density , obeying eigenstate thermalization hypothesis (ETH) for driven systems~\cite{PhysRevE.90.012110}   -- with the exception of a minority of quantum scars, and these scars display time translation symmetry breaking, so that they support persistent period doubling for some class of initial states. This scenario has been observed in a periodically kicked PXP model~\cite{Maskara_2021,PhysRevB.108.205129}, driven interacting spinless fermions~\cite{PhysRevB.102.224309}, and driven interacting bosons on a lattice~\cite{PhysRevLett.129.133001}. In this work we find this phenomenon also in a kicked Ising spin chain with power law interactions. In order to highlight it we consider the properties of the Floquet states and we find a fraction of Floquet states showing both $\pi$-spectral pairing and long-range order. Instead of probing long-range order using the usual correlator considered in Ref.~\cite{PhysRevLett.117.090402}, we diagonalize the magnetization operator in each $\pi$-spectral paired doublet. The Floquet states in each doublet are cat states, so the magnetization has vanishing expectation on them. Therefore, one must consider an appropriate superposition of states in the doublet in order to find an extensive nonvanishing magnetization value and probe long-range order. This approach has already been used to find symmetry breaking in excited states of undriven Hamiltonians~\cite{gómez2025quantumthermalizationmechanismemergence}.

 We find a clear tendency of Floquet states in doublets with a large normalized magnetization eigenvalue to have a small value of the half-system entanglement entropy, much smaller than the Page value displayed by thermal states. We quantify the number of nonthermal scar-like Floquet states by looking at the number of states in doublets with a normalized magnetization eigenvalue larger than a given threshold. We find that this number increases exponentially with the system size. Thus, although these states are a minority in the Hilbert space -- they become a vanishing fraction in the thermodynamic limit -- they are indeed quite relevant and have significant effects on the dynamics, similarly to the undriven case studied in Ref.~\cite{kv2w-mk4t}. The result is that --although the majority of the Floquet states are thermal, as highlighted by the level-spacing distribution -- there are many initial states that lead to persistent period doubling.

We find that taking as initial states some classical spin configurations that have huge overlap with the time-translation symmetry-breaking Floquet states one can see persistent period doubling oscillations, and otherwise there is not such a phenomenon. The initial configurations leading to long-lasting period-doubling oscillations are those where the maximum of the distribution of the (normalized) magnetization eigenvalue over the doublets weighted by the overlap with the initial state is nonvanishing. We consider as initial states on one hand classical configurations with domain walls and on the other configurations where the spins are uniformly tilted. In the latter case we can perform a finite-size scaling. We find that in the presence of persistent period doubling the relevant $\pi$-shifted gap decreases exponentially with the system size. This suggests that in this case period doubling oscillations last for a time exponential in the system size, as typically occurs in Floquet time crystals~\cite{PhysRevLett.117.090402,PhysRevB.109.174310,Surace_2019,PhysRevB.95.214307}.

The paper is organized as follows. In Sec.~\ref{mod:sec} we describe the model and the time-translation symmetry breaking physics. In Sec.~\ref{res:sec} we discuss our numerical results and show that, for appropriate initial conditions, time-translation symmetry breaking physics appears for different values of the power-law decay exponent, also beyond the range where mean-field physics is valid in the thermodynamic limit. In Sec. \ref{sec:tilted} we consider tilted initial states, which allow for clean finite-size scaling analyses. Finally, in Sec.~\ref{sec:conclusions} we draw our conclusions.

\section{The model and time-translation symmetry-breaking physics}\label{mod:sec}
\subsection{The model}
We consider the long-range 1D Ising model with both transverse and longitudinal fields,
\begin{equation}
  \label{eq:hamiltonian}
 \hat{H} = -\frac{J}{\mathcal{K}(N,\alpha)} \sum_{i\neq j} \frac{1}{D_{ij}^{\alpha}} \, \hat{\sigma}^z_i \hat{\sigma}^z_j + h_z \sum_i \hat{\sigma}^z_i + h_x \sum_i \hat{\sigma}^x_i.
\end{equation}
In this equation, $J$ is a coupling constant, $h_{x}$ is a transverse magnetic field along the x-axis, $h_{z}$ is a longitudinal magnetic field along the $z$-axis, and $\alpha$ accounts for the power-law interaction between spins. The term $\mathcal{K}(N,\alpha)$ is the Kac factor \cite{Kac1963}, which ensures extensivity of the energy for $\alpha \leq 1$, and the distance is defined as $D_{ij}\equiv\min\{|i-j|,N-|i-j|\}$, in order to implement periodic boundary conditions~\cite{Liu2019}. In this geometry, the Hamiltonian is symmetric under
the full set of transformations of the dihedral group $D_N$. Choosing initial states consistent with this symmetry, dynamics is restricted to the subspace with zero momentum and positive inversion parity, reducing the size of the relevant Hamiltonian matrix, allowing thereby to numerically reach larger system sizes. Further details on the symmetries of Eq.~\eqref{eq:hamiltonian} can be found in Appendix \ref{app:symmetries}.

We apply a periodically-kicked time evolution that consists of the two steps described below:

\textbullet\, A unitary evolution given by the Hamiltonian in Eq. \eqref{eq:hamiltonian} acting during an interval $\tau$,
  \begin{equation}
    \hat{U}_H = \textrm{e}^{-i \hat{H}\tau},
  \end{equation}
  where we assume $\hbar=1$.

  \textbullet\, A {\em kick} given by the unitary operator
  \begin{equation}
    \hat{U}_K = \textrm{e}^{i \left( \frac{\pi}{2} + \epsilon \right) \sum_i \hat{\sigma}^x_i},
  \end{equation}
  where $\epsilon\in\mathbb{R}$ is a free parameter.

Then, the unitary operator $\hat{U}$ is periodic with period $\tau$:
\begin{equation}
  \hat{U}(\tau) = \hat{U}_{K} \hat{U}_H \Rightarrow \hat{U}(n\tau) = \left[ \hat{U}_{K} \hat{U}_H \right]^n.
  \end{equation}

If $\epsilon=h_z=0$, then $[\hat{H},\hat{U}_{K}]=0$ and therefore the kick does not dissipate energy. 

Let us discuss the properties of the undriven Hamiltonian Eq.~\eqref{eq:hamiltonian} relevant for our analysis. For our purposes, three main physical aspects of this model have been previously studied:

\begin{enumerate}
\item For $h_{z}=0$, Eq. \eqref{eq:hamiltonian} exhibits a thermal phase transition when $0<\alpha<2$. In this context, ferromagnetic equilibrium states, which break the global Hamiltonian symmetry, have been previously observed in Refs. \cite{GonzalezLazo2021,Corps2022,Neyenhuis2017,Halimeh2017,Zunkovic2018,Piccitto2019,Ranabhat2022}.

\item This Hamiltonian is also known to support the phenomenon of domain wall confinement. This has been observed in the parameter range where a thermal phase transition is present, for $0<\alpha<2$, for $\alpha\gtrsim 2$ but not too large and $h_{z}=0$, and in the nearest-neighbors case $\alpha\to\infty$ and $h_{x}\neq 0$. Yet, this theory is usually applied to polarized initial states (with no magnetic domains) and to states with a single magnetic domain. For $h_{z}=0$, these observations were reported in Refs. \cite{Hauke2013,Lerose2019,Liu2019,Tan2021,Verdel2020,Vovrosh2022}. The relationship between domain wall confinement and time crystals has been previously considered in Ref. \cite{Collura2022}. For $h_{z}\neq 0$, domain wall confinement was studied in Refs. \cite{Collura2022,Vovrosh2022,Verdel2023,Lerose2020}. A theoretical model for domain wall confinement of states with a single magnetic domain was proposed in Ref. \cite{Liu2019}. In this paper, however, we consider initial states with a higher number of magnetic domains, as well as tilted states that are not described by this framework.

\item Recently, it was shown in Ref. \cite{kv2w-mk4t} that the one-dimensional transverse-field Ising model with long-range interactions and $h_{z}=0$ supports a set of ferromagnetic scarred eigenstates embedded in a sea of conventional paramagnetic states. These scarred states are nonthermal and therefore violate the eigenstate thermalization hypothesis, as they exhibit ferromagnetic order despite being surrounded by spectrally nearby paramagnetic eigenstates. Their existence and robustness depend on the interaction-range exponent $\alpha$: as $\alpha$ increases, the scarred structure progressively weakens, although it can persist even for $\alpha>2$. For $\alpha<1$, the origin of these scars can be traced back to the universality class of the fully connected limit. Here, we exploit these scarred states in a Floquet setting and show that they give rise to robust time-translation symmetry breaking, thereby realizing a time-crystalline phase. Remarkably, this phenomenon survives even when the system is globally chaotic, in the sense that its quasienergy spectrum follows the predictions of random-matrix theory. 
\end{enumerate}

\subsection{Floquet time crystal physics}

Let us consider the eigenstates (Floquet states) and eigenvalues (Floquet quasienergies) of the Floquet operator~\cite{PhysRevA.7.2203,PhysRev.138.B979}:
\begin{equation}
  \hat{U}(\tau) \ket{\phi_p} = \textrm{e}^{-i \alpha_p\tau} \ket{\phi_p}, \; \; \alpha_p \in \mathbb{R},\, \, \forall p.
\end{equation}
Then, after $n$ periods, the initial condition
\begin{equation}
  \ket{\Psi(0)} = \sum_p c_p \ket{\phi_p},
\end{equation}
where $\sum_{p}|c_{p}|^{2}=1$, becomes
\begin{equation}
  \ket{\Psi(n\tau)} = \sum_p c_p \textrm{e}^{-i n \alpha_p\tau} \ket{\phi_p}.
\end{equation}
Therefore, the expectation value of any observable, $\hat{M}$, is
\begin{equation}
  \langle \hat{M}(n) \rangle = \sum_{p,q} c_q^* c_p \textrm{e}^{i n\tau (\alpha_q - \alpha_p)} \bra{\phi_q} \hat{M} \ket{\phi_p}.
\end{equation}
To better understand the meaning of this equation, let us introduce the short writing $M_{pq} \equiv \bra{\phi_q} \hat{M} \ket{\phi_p}$ and split it in three terms:
\begin{equation}\label{Mnt:eqn}
  \begin{split}
    \langle \hat{M}(n) \rangle &= \sum_{p} |c_p|^2  M_{pp}    +  \Big(\sum_{\substack{p,q\\ \alpha_p - \alpha_q = \pi/\tau \\ \textrm{ mod} (2\pi) }} c_q^* c_p \textrm{e}^{-i n \pi} M_{qp} + \textrm{H.c.}\Big) \\ &+ \Big(\sum_{\substack{p,q\\ \alpha_p - \alpha_q \neq \pi \\ \textrm{ mod} (2\pi) }} c_q^* c_p \textrm{e}^{-i n\tau (\alpha_p - \alpha_q)} M_{qp}+\textrm{H.c.}\Big)
    \end{split}
\end{equation}
The key term for a Floquet time crystal is the second one: If it exists, it induces a coherent oscillation with period $2\tau$. In such a case, the first term represents the average value around which this wave oscillates, and the third term generates some imperfections in the time crystal. 

Thus, in order to observe period-doubling oscillations, the second term must be nonnegligible.
We formulate this requirement in the following way:

\textbullet\, For each Floquet eigenstate, $\ket{\phi_p}$, we choose a {\em partner}, $\ket{\phi_q}$, that minimizes the $\pi$-shifted gap $\Delta_{pq}^{\pi} = \left| |\alpha_p - \alpha_q| - \pi/\tau\right|$.~\footnote{Consider that the quasienergies lie in the first Floquet Brillouin zone, $-\pi/\tau\leq \alpha_p\leq \pi/\tau$.} In this way we identify the $\pi$-shifted doublets that are essential for the period-doubling behavior, and the gap in each doublet is given by
\begin{equation}
  \Delta_p^\pi \equiv \min_q\left| |\alpha_p - \alpha_q| - \pi/\tau\right|\,.
\end{equation}
 Notice that if the $\Delta_p^\pi$ are not rigorously vanishing -- as in the second term on the right side of Eq.~\eqref{Mnt:eqn} -- the period doubling oscillations are dephased away after a time scale $\sim 1/\Delta_{p}^{\pi}$, so it is very important that for the doublets relevant for the dynamics $\Delta_{p}^{\pi}\ll 1$. In Floquet time crystals the logarithmic average over Floquet states of this quantity vanishes in the thermodynamic limit when compared to the logarithmic average gap between nearby Floquet states~\cite{PhysRevB.94.085112}.

\textbullet\, Floquet states in a $\pi$-spectral paired doublet must be long-range correlated, i.e., they must be cat-state superpositions of macroscopically magnetized states. In order to highlight this property, for each Floquet state doublet we have identified minimizing $\Delta_{pq}^\pi$, we build the subspace, $\mathcal{E}_p$, spanned by $\lbrace \ket{\phi_p}, \ket{\phi_q} \rbrace$. In each $2\times 2$ subspace we diagonalize the operator 
\begin{equation}\label{Mz:eqn}
  \hat{J}_z\equiv\sum_{j=1}^L \hat{\sigma}_j^z
\end{equation}
 and obtain the eigenvalues, $\Lambda_{p,+}$ and $\Lambda_{p,-}$, and the eigenvectors, $\ket{\widetilde{\phi}_{p,+}}$ and $\ket{\widetilde{\phi}_{p,-}}$. If these eigenvalues are extensive we know that the doublet shows long-range correlation.

Notice that the first condition corresponds to $\pi$-spectral pairing and the second one to the states in a doublet being cat-state superposition of macroscopically magnetized states. Together, these conditions correspond to the states in the considered doublet displaying time-translation symmetry breaking. We can connect the period doubling oscillations occurring for a given initial state with the time-translation symmetry breaking properties of the Floquet states with relevant overlap with that state by quantitatively checking the two properties listed above.

\textbullet\, In order to get a time crystal, the $\Delta_{p}^\pi$ involved into the dynamics for a given initial state should be small enough that there is no dephasing of the period-doubling oscillations in the thermodynamic limit, in a sense that we are going to clarify. For that purpose we define the logarithmic $\pi$-shifted gap overlap-weighted distribution
\begin{equation}\label{Pd:eqn}
    P(\log_{10}\Delta^\pi) = \sum_p |\braket{\Psi(0)|\phi_p}|^2\delta(\log_{10}\Delta^\pi -\log_{10}\Delta_p^\pi)\,.
\end{equation}
One can easily check that this distribution is normalized to one. Let us consider the average of this distribution
\begin{equation}\label{lode:eqn}
  \braket{\log_{10}\Delta^\pi} \equiv \sum_p |\braket{\Psi(0)|\phi_p}|^2\log_{10}\Delta_p^\pi
\end{equation}  
whose exponential provides the inverse lifetime of the period-doubling oscillations. (After this timescale the period-doubling oscillations get dephased away). If $\braket{\log_{10}\Delta^\pi}\to-\infty$ for $L\to\infty$, then in this limit the period-doubling oscillations undergo no dephasing, because their lifetime diverges and the system behaves as a time crystal. In Sec.~\ref{sec:tilted} we are going to show that in a class of initial states where finite-size scaling is possible, $\braket{\log_{10}\Delta^\pi}$ linearly decreases with system size, and extrapolating to $L\to\infty$ one gets that the no-dephasing condition $\braket{\log_{10}\Delta^\pi}\stackrel{\tiny L\to\infty}{\longrightarrow}-\infty$ is fulfilled.

\textbullet\, Defining the normalized eigenvalues
\begin{equation}
    \lambda_{p,\pm} \equiv \Lambda_{p,\pm} / N,
\end{equation}
one must have $|\lambda_{p,\pm}| \nrightarrow 0$ in the thermodynamic limit. In order to focus on the $\lambda_{p,\pm}$ most relevant for the dynamics we define -- similarly to Eq.~\eqref{Pd:eqn} -- the normalized-eigenvalue overlap-weighted distribution
\begin{equation}\label{Pl:eqn}
    P(\lambda) = \frac12\sum_p |\braket{\Psi(0)|\phi_p}|^2[\delta(\lambda-\lambda_{p,+})+\delta(\lambda-\lambda_{p,-})]\,,
\end{equation}
coarse grain it, and study its mode (the value of $\lambda$ corresponding to the maximum of the distribution). As we show in the next section, $\lambda_{\rm max}$ -- the value of $\lambda$ corresponding to this mode -- is the one most relevant for the dynamics.  We call  the coarse-grained versions of the distributions as $P_{\rm c.g.}$, and from now on we fix $\tau=1$.
~\footnote{Beyond these two conditions, there is a third one: The non-diagonal elements between different doublets must be small compared to the ones inside the doublets so that the second term in Eq.~\eqref{Mnt:eqn} dominates over the third one. We do not check it here because the other two are enough for predicting if a given initial state will give rise to persistent period doubling.}

In the next section we will show how these conditions are satisfied for only a fraction of the Floquet states inside a bulk of thermal Floquet states, giving rise to a period-doubling associated to a weak ergodicity breaking phenomenon akin to quantum scars.

\section{Initial states with magnetic domains}\label{res:sec}

\begin{center}
\begin{figure*}[t]
  \begin{tabular}{cc}
  (a) & (b) \\
  \includegraphics[width=66mm]{{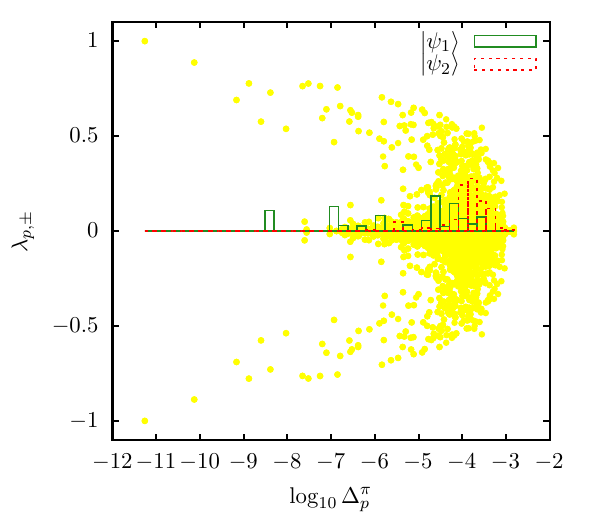}} &
  \includegraphics[width=66mm]{{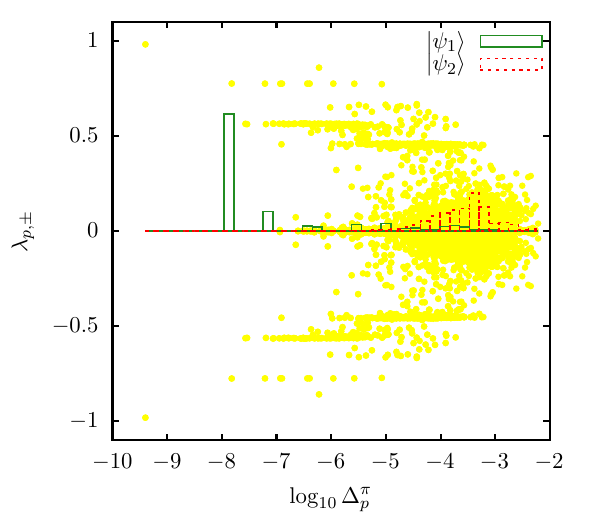}}\\
  (c) & (d)\\
  \includegraphics[width=66mm]{{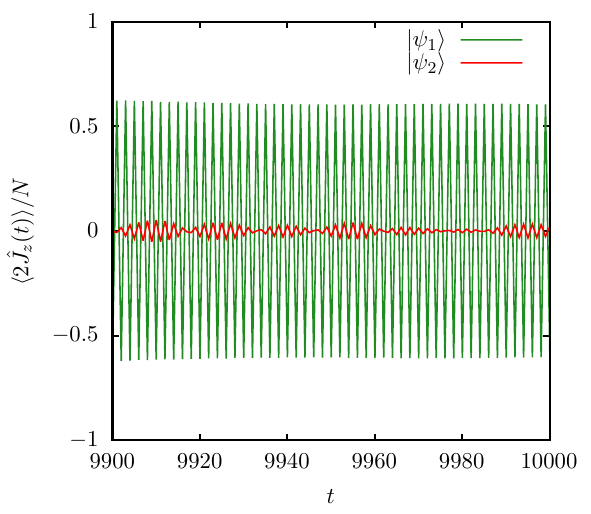}} &
  \includegraphics[width=66mm]{{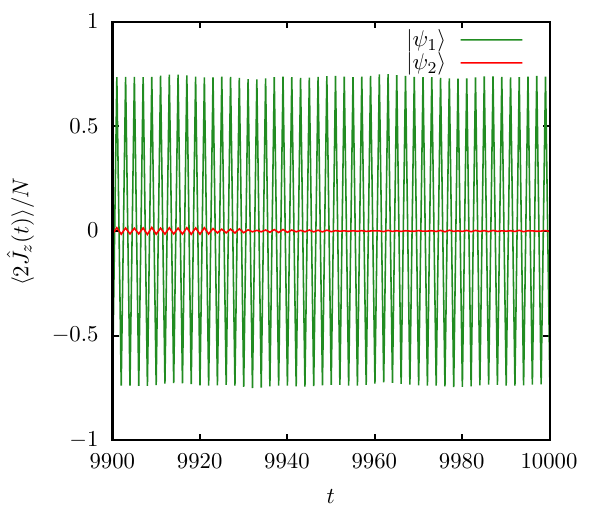}}\\
  (e) & (f) \\
 \includegraphics[width=66mm]{{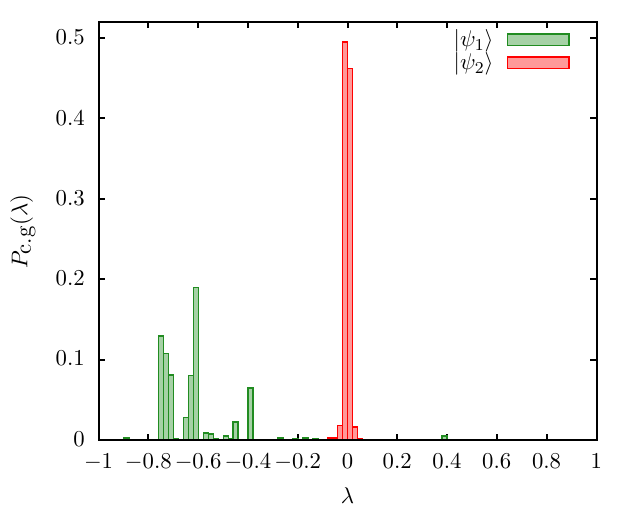}} &
 \includegraphics[width=66mm]{{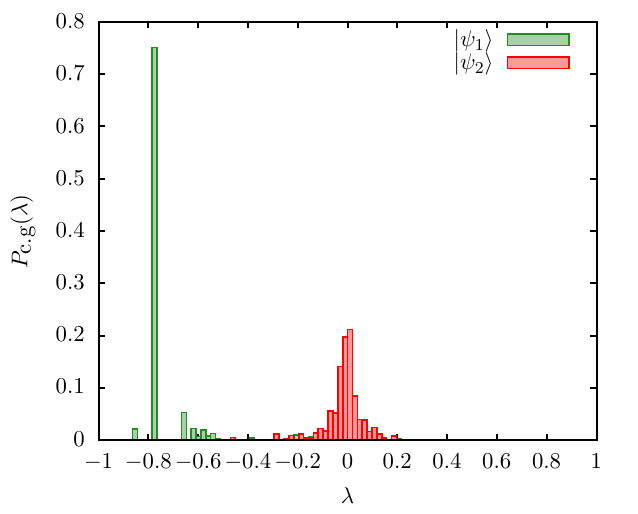}}
\end{tabular}
\caption{(a)-(b) Scatter plots where each yellow point corresponds to a Floquet-state doublet with normalized magnetization eigenvalues on the vertical axis and the logarithm of the $\pi$-shifted gap on the horizontal one. On the same plots we plot $P_{\rm c.g.}(\log_{10}\Delta^\pi)$ [coarse-grained version of Eq.~\eqref{Pd:eqn}] for the two chosen initial states $\ket{\psi_{1}},\,\ket{\psi_2}$. $\ket{\psi_1}$ has overlap with doublets with a $\pi$-shifted gap many orders of magnitude smaller than $\ket{\psi_2}$. (c)-(d) Time evolution of the $z$ magnetization with the two chosen initial states. The state $\ket{\psi_1}$ -- the one with smaller $\pi$-shifted gap -- provides period-doubling , while the other doesn't. (e)-(f) $P_{\rm c.g}(\lambda)$ [coarse-grained version of Eq.~\eqref{Pl:eqn}]. $\ket{\psi_1}$ has overlap with doublets with a normalized eigenvalue significantly different from zero. Numerical parameters for the first column: $J=1,\,h_x=0.1,\,h_z=0.2,\,\alpha=2.2,\,\epsilon=0,\,N=19$. Numerical parameters for the second column: $J=1,\,h_x=0.25,\,h_z=0.0,\,\alpha=0.6,\,\epsilon=0.1,\,N=19$.}\label{figZ:fig}
\end{figure*}
\end{center}
In Fig.~\ref{figZ:fig} we consider two cases, one with nonvanishing longitudinal field $h_z$ and $\epsilon=0$ (left panels) and one with vanishing longitudinal field and nonvanishing $\epsilon$ (right panels). Both situations are nontrivial because in both cases the kicking operator does not commute with the Hamiltonian. In Fig.~\ref{figZ:fig}(a)-(b) we show scatter plots where each point corresponds to a doublet of Floquet states. On the horizontal axis we show the logarithm of the $\pi$-shifted gap $\Delta_p^{\pi}$, and in the vertical axis the normalized magnetization eigenvalues. We see that doublets with larger eigenvalue show a $\pi$-shifted gap of smaller order of magnitude, marking the strict relation between long-range order and $\pi$-spectral pairing mentioned above.

In the same panels, we consider also two possible initial states that have the form of classical spin configurations with magnetic walls. We take as initial states
\begin{align}\label{states:eqn}
  \ket{\psi_1}&=\ket{\uparrow\downarrow\downarrow\downarrow\downarrow\downarrow\downarrow\downarrow\downarrow\downarrow\downarrow\uparrow\downarrow\downarrow\downarrow\downarrow\downarrow\downarrow\downarrow\nonumber}\\
  \ket{\psi_2}&=\ket{\uparrow\uparrow\uparrow\uparrow\downarrow\downarrow\downarrow\downarrow\downarrow\downarrow\uparrow\uparrow\uparrow\downarrow\downarrow\downarrow\downarrow\downarrow\downarrow}\,.
\end{align}
 Both states have the same number of domain walls but behave differently considering their overlap with the Floquet doublets. We can see this in Fig.~\ref{figZ:fig}(a)-(b) where we also plot $P_{\rm c.g.}(\log_{10}\Delta^\pi)$ [the coarse-grained version of Eq.~\eqref{Pd:eqn}]. In order to visualize this distribution, we need to coarse-grain it, by defining a binning \emph{i.~e.} a discretization of the horizontal axis. We find that $\ket{\psi_1}$ has a significant overlap with doublets where the $\pi$-shifted gap is very small (order $10^{-8}$) while $\ket{\psi_2}$ has predominantly contributions from the bulk of states with larger $\pi$-shifted gap. This is true for both choices of parameters in panels (a) and (b). 
 
 In a more quantitative way, this difference reflects in the average of $\log_{10}\Delta^\pi$ [see Eq.~\eqref{lode:eqn}]. In panel (a) $\ket{\psi_1}$ gives rise to  $\braket{\log_{10}\Delta^\pi}=-5.32042$, while $\ket{\psi_2}$ corresponds to $\braket{\log_{10}\Delta^\pi}=-3.9190$. In panel (b) the difference is even more marked, with $\ket{\psi_1}$ giving rise to $\braket{\log_{10}\Delta^\pi}=-6.9027$ and $\ket{\psi_2}$ corresponding to $\braket{\log_{10}\Delta^\pi}=-3.6051$. In the panel-(b) case, we expect therefore that with $\ket{\psi_1}$ as initial state the period-doubling oscillations last for a lifetime three orders of magnitude than with $\ket{\psi_2}$.

\begin{figure*}
\begin{center}
  \begin{tabular}{cc}
  (a) & (b) \\
  \includegraphics[width=67mm]{{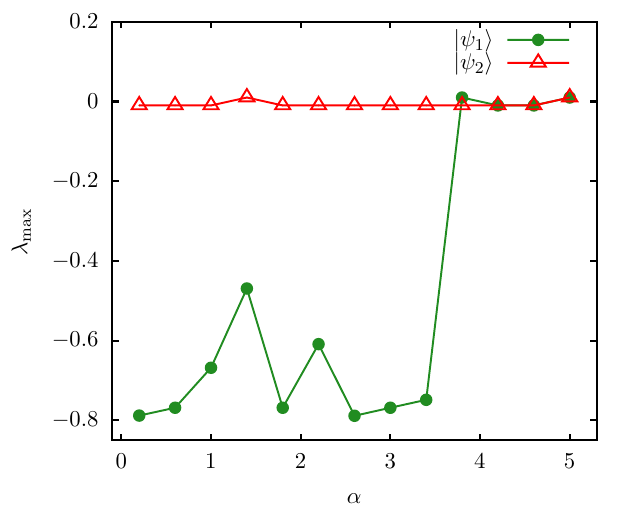}} &
  \includegraphics[width=67mm]{{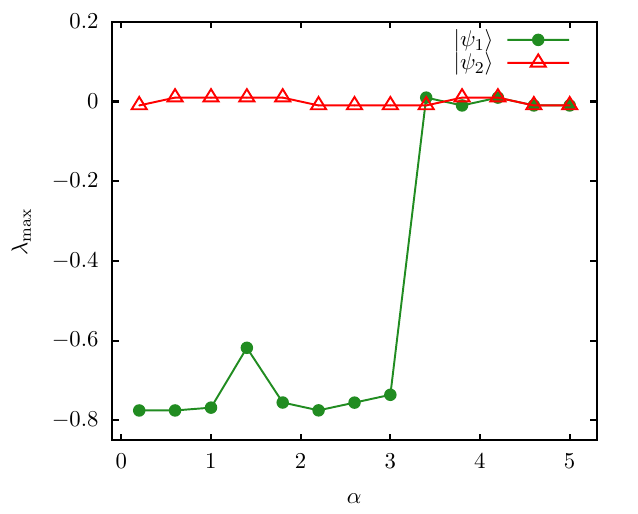}}\\
  \end{tabular}
  \caption{Mode of the coarse-grained normalized-eigenvalue distribution for the two chosen initial states and two different parameter choice. Numerical parameters for panel (a): $J=1,\,h_x=0.1,\,h_z=0.2,\,\epsilon=0,\,N=19$. Numerical parameters for panel (b): $J=1,\,h_x=0.25,\,h_z=0.0,\,\epsilon=0.1,\,N=19$. Bin width in the coarse graining of Eq.~\eqref{Pl:eqn} is $0.02$}\label{figmax:fig}
  \end{center}
\end{figure*}
The contrast between the two initial states is even stronger if we look at the dynamics. Considering the expectation value of the $z$-magnetization Eq.~\eqref{Mz:eqn} versus time with the two considered initial states we see a striking difference [Fig.~\ref{figZ:fig}(c)-(d)]. For $\ket{\psi_1}$ we see persistent period doubling oscillations, which presumably last for a time of order $t^*\sim \exp\left(-\braket{\log_{10}\Delta^\pi}\right)$, so $t^*\sim10^{5}$ in panel (a) and $t^*\sim10^{7}$ in panel (b). By contrast, initializing at $\ket{\psi_2}$ we do not observe any such persistent period-doubling oscillations, but just some small irregular wigglings.

In Ref. \cite{kv2w-mk4t} initial states analogous to those in Eq. \eqref{states:eqn} were considered to reveal ferromagnetic scars in the case $h_{z}=0$. According to the results of that work, in the case of $h_{z}=0$ and no kick, $\ket{\psi_{1}}$ evolves towards a ferromagnetic equilibrium state, while $\ket{\psi_{2}}$ gives rise to a paramagnetic equilibrium state, even for values of $\alpha\gtrsim 2$. Here, we observe that this behavior is robust, as it generates time crystals even in the presence of a kick that does not commute with the Hamiltonian for two completely different choices of parameters: $h_{z}\neq 0$ and $\epsilon\neq0$.

\begin{figure*}
\begin{center}
  \begin{tabular}{cc}
  (a) & (b) \\
  \includegraphics[width=70mm]{{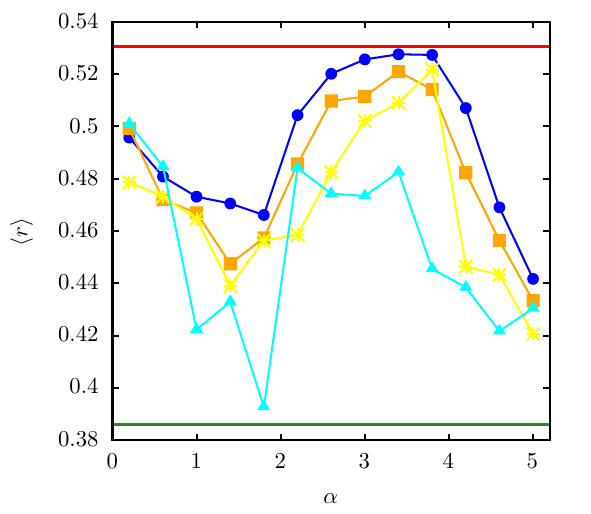}} &
  \includegraphics[width=70mm]{{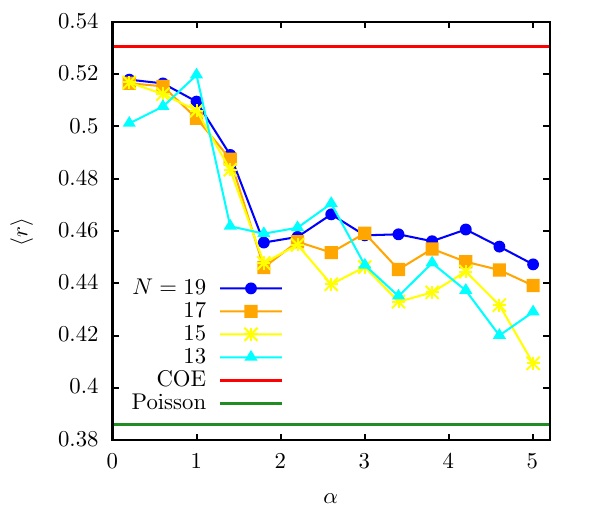}}\\
  \end{tabular}
  \caption{Average level spacing ratio, $\langle r\rangle$, versus $\alpha$. Numerical parameters for panel (a): $J=1,\,h_x=0.1,\,h_z=0.2,\,\epsilon=0,\,N=19$. Numerical parameters for panel (b): $J=1,\,h_x=0.25,\,h_z=0.0,\,\epsilon=0.1,\,N=19$. The COE and Poisson theoretical values are plotted with red and green horizontal lines. For (b), the Hamiltonian commutes with the parity operator Eq. \eqref{eq:parity}; therefore, the ratio reported here is the average ratio of the positive and negative parity sectors. }\label{COE:fig}
  \end{center}
\end{figure*}

We can better understand this phenomenon by looking at $P_{\rm c.g.}(\lambda)$ [the coarse grained version of Eq.~\eqref{Pl:eqn} -- see Fig.~\ref{figZ:fig}(e)-(f)]. We see that for $\ket{\psi_1}$ there is  a clear maximum at a value different from 0 and of order one. The absolute value of this maximum point coincides with the amplitude of the corresponding period-doubling oscillations and we have checked that this is generally true. For $\ket{\psi_2}$ this maximum is at 0 and consistently there are no period-doubling oscillations.

\begin{figure*}
\begin{center}
  \begin{tabular}{cc}
  (a) & (b) \\
  \includegraphics[width=80mm]{{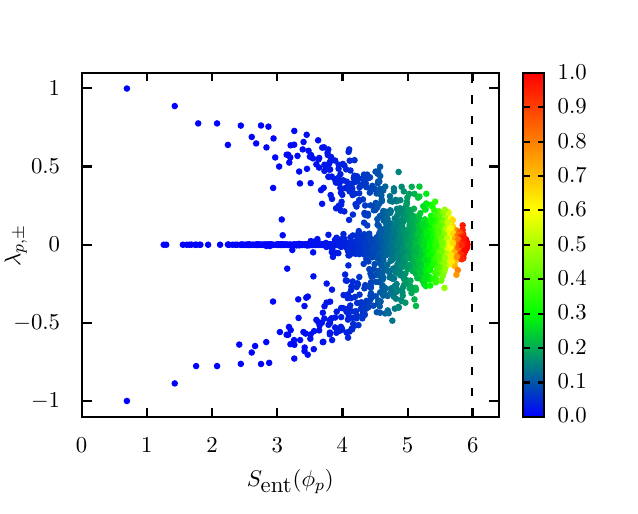}} &
  \hspace{-0.7cm}\includegraphics[width=80mm]{{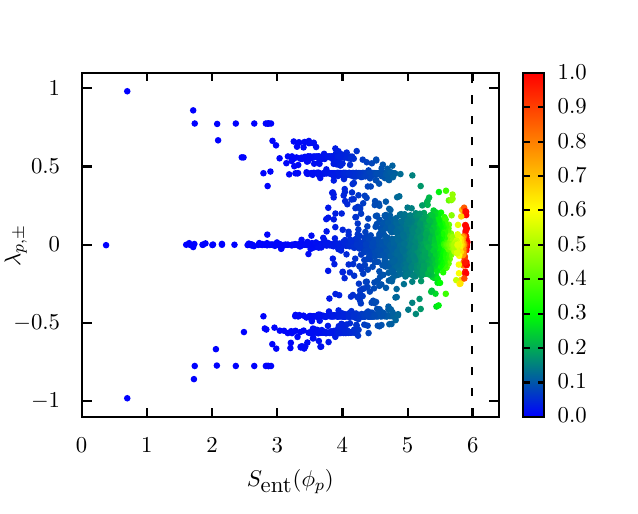}}\\
  \end{tabular}
  \caption{Scatter plots of the Floquet doublets with half-chain entanglement entropy of the states in the doublet on the horizontal axis and normalized magnetization eigenstates on the vertical one. Numerical parameters for panel (a): $J=1,\,h_x=0.1,\,h_z=0.2,\,\alpha=2.2,\,\epsilon=0,\,N=19$. Numerical parameters for panel (b): $J=1,\,h_x=0.25,\,h_z=0.0,\,\alpha=0.6,\,\epsilon=0.1,\,N=19$. The Page value for a fully ergodic quantum state is represented by a black vertical line.}\label{entro:fig}
  \end{center}
\end{figure*}

The ones shown in Fig.~\ref{figZ:fig} are particular cases for a specific choice of $\alpha$. We have also checked that this picture is valid for different values of $\alpha$, as we discuss below.  

\begin{figure*}
\begin{center}
  \begin{tabular}{cc}
  (a) & (b) \\
  \includegraphics[width=70mm]{{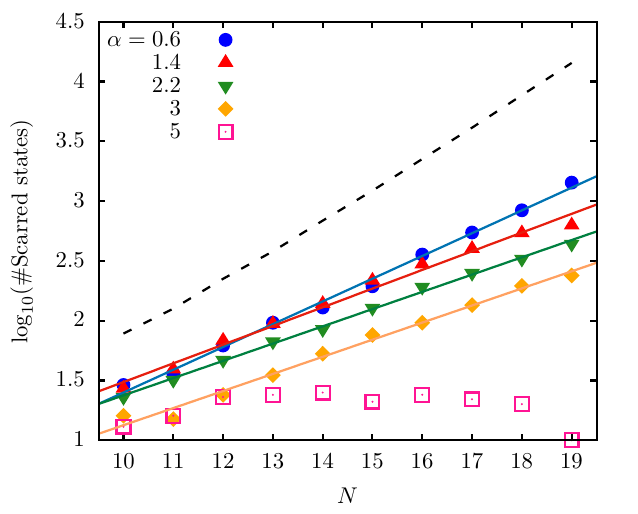}} &
  \includegraphics[width=70mm]{{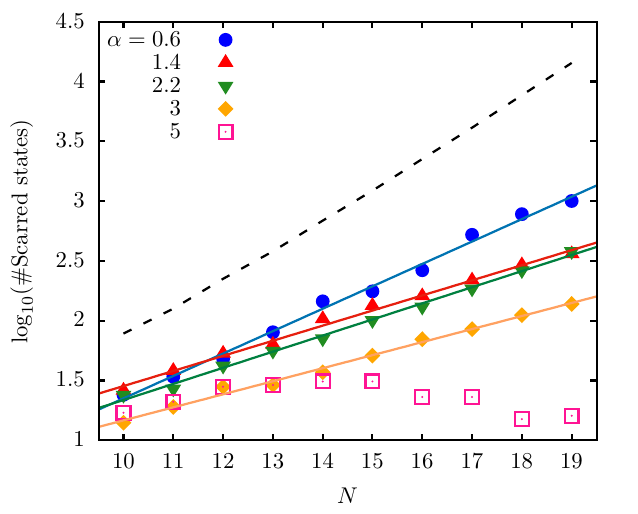}}\\
  \end{tabular}
  \caption{Logarithm of the number of long-range ordered Floquet eigenstates versus $N$ for different parameter choices. Notice the linear increase for $\alpha < 5$, as highlighted by the linear least-square fits. We take as long-range ordered Floquet eigenstates those with absolute value of the normalized magnetization eigenvalue larger that 0.2. (We have checked that taking a different threshold does not change results qualitatively.) Numerical parameters for panel (a): $J=1,\,h_x=0.1,\,h_z=0.2,\,\epsilon=0,\,N=19$. Numerical parameters for panel (b): $J=1,\,h_x=0.25,\,h_z=0.0,\,\epsilon=0.1,\,N=19$. The total dimension of the Hilbert subspace relevant for the dynamics for each $N$ is plotted with a dashed line. }\label{slope:fig}
  \end{center}
\end{figure*}

We highlight this feature by looking at the maximum point $\lambda_{\rm max}$ of the coarse-grained distribution Eq.~\eqref{Pl:eqn}, two examples of which are shown in Fig.~\ref{figZ:fig}(e)-(f). We plot this maximum point versus $\alpha$ in Fig.~\ref{figmax:fig} for the two initial states and the two choices of $\epsilon$ and $h_z$ that we have discussed above. The contrast between the two initial states is striking. While $\ket{\psi_1}$ displays a maximum with an absolute value on the order of 1 for $\alpha\leq 3$ (that is to say, well beyond the mean-field regime), the maximum of $\ket{\psi_2}$ is always very close to 0. Thus, the first initial state has a significant overlap with Floquet states exhibiting long-range order.

We remark that, in contrast to the disordered case where all the Floquet spectrum displays $\pi$ spectral pairing, here only a part of the Floquet eigenstates shows this property, while the majority of them are thermal ones, a behavior resembling quantum scars. We highlight this phenomenon by looking at the average level spacing ratio~\cite{PhysRevB.75.155111}, defined as
\begin{equation}
  \braket{r} = \frac{1}{\mathcal{N}-2}\sum_{p=1}^{\mathcal{N}-2}\frac{\min(\alpha_{p+1}-\alpha_p,\alpha_{p+2}-\alpha_{p+1})}{\max(\alpha_{p+1}-\alpha_p,\alpha_{p+2}-\alpha_{p+1})}\,,
\end{equation}
where the Floquet quasienergies $\alpha_p$ are in increasing order and are restricted to the largest irreducible subspace under the symmetries (the one with zero momentum and positive inversion parity -- see Appendix~\ref{app:symmetries}), whose dimension is $\mathcal{N}$. If this quantity has a value $\braket{r}\simeq 0.38$ then the dynamics is regular and thermalization is hindered by an extensive number of local integrals of motion. In contrast, when it is $\braket{r}\simeq 0.52$ the spectrum is similar to that of a circular orthogonal ensemble (COE) random matrix and the dynamics leads to thermalization~\cite{PhysRevLett.110.084101}.
We plot $\braket{r}$ versus $\alpha$ Fig.~\ref{COE:fig} and we see that it is always markedly different from the Poisson value corresponding to perfect regular dynamics. This shows that at least part of the spectrum is composed of thermal states.

To better understand the role of the nonthermal states, in Fig.~\ref{entro:fig} we show two examples of scatter plots such that each point corresponds to a doublet. On the horizontal axis we plot the half-chain entanglement entropy\footnote{The half-chain entanglement entropy is defined as the von Neumann entropy of the density matrix obtained tracing out half of the system.} of the Floquet states belonging to that doublet, and on the vertical axis the normalized magnetization eigenvalues for that doublet. We see that long-range ordered states -- those with magnetization eigenvalues more distant from 0 -- are also the ones with smaller entanglement entropy. They are a minority of the Floquet spectrum, and most of the Floquet states are near the Page value~\cite{PhysRevLett.71.1291} marked as a vertical line in the plot, as highlighted by the color scale marking the cumulative density of the data. The Page value (evaluated as in~\cite{PhysRevB.104.094309}) corresponds to the case of a random state and states that are locally thermal with infinite temperature are expected to have this value of the half-chain entanglement entropy~\cite{PhysRevB.101.064302}.

The nonthermal states are a minority of the Floquet spectrum, but they are not so few. In order to highlight this, we plot in Fig.~\ref{slope:fig} the logarithm of the number of eigenstates in a doublet with at least one  normalized magnetization eigenvalue whose absolute value lies above a given threshold versus the number of sites $N$. We see a linear increase with a slope smaller thane one corresponding to the Hilbert-space dimension. This increase is quite robust and disappears only for very large $\alpha$. In particular the increase is clearly visible also for $\alpha=3$ that is quite above the threshold $\alpha=1$ below which a mean-field behavior in the thermodynamic limit is expected~\cite{kv2w-mk4t}. Thus, these eigenstates are a vanishing fraction in the thermodynamic limit, but their number is still exponential in $N$, and can deeply affect the dynamics of experimentally relevant (and common) initial states, like $\ket{\psi_1}$ in Eq.~\ref{states:eqn}, providing long-lasting period-doubling oscillations. 

As a last remark, notice that the low-entanglement points  in Fig.~\ref{entro:fig} are organized along a curve, similarly to what happens in nondriven quantum scarred systems when the entanglement entropy of the eigenstates is plotted against their energy (see for instance Fig.~4 of Ref.~\cite{You_2022}). Performing here a similar plot against the quasienergies would not provide such a well organized curve due to the folding of the quasienergy spectrum.

\section{Finite-size scaling: tilted states}\label{sec:tilted}

\begin{figure*}
\begin{center}
  \begin{tabular}{cc}
  (a) & (b) \\
  \includegraphics[width=68mm]{{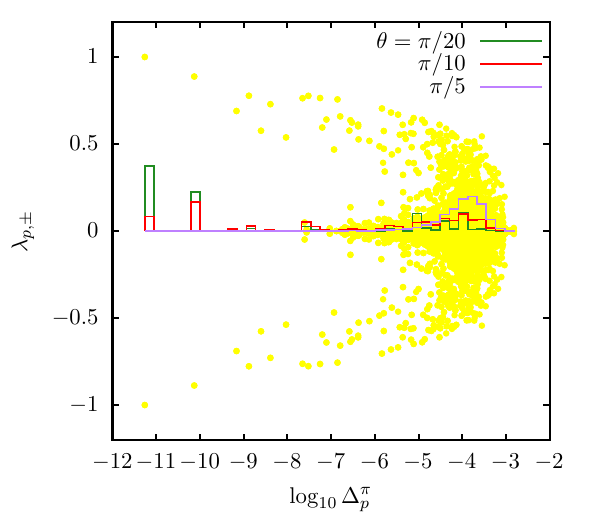}} &
  \includegraphics[width=68mm]{{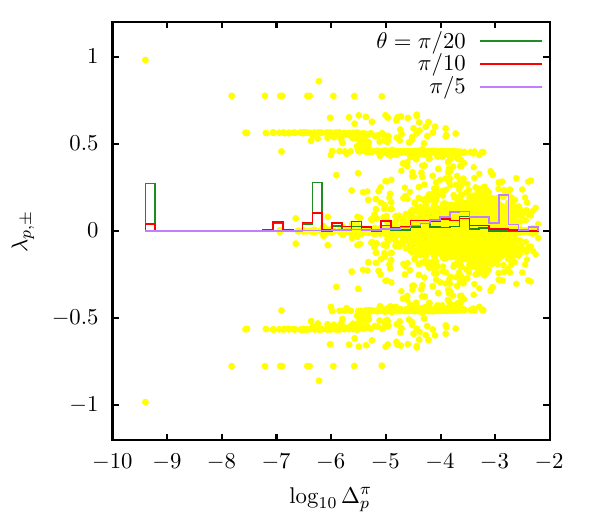}}\\
  (c) & (d)\\
  \includegraphics[width=68mm]{{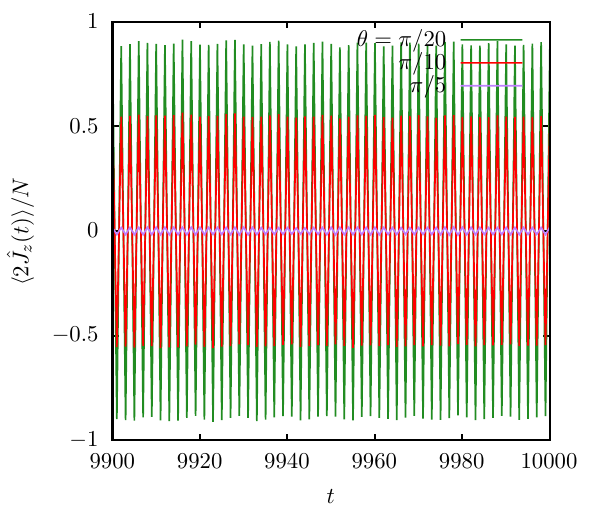}} &
  \includegraphics[width=68mm]{{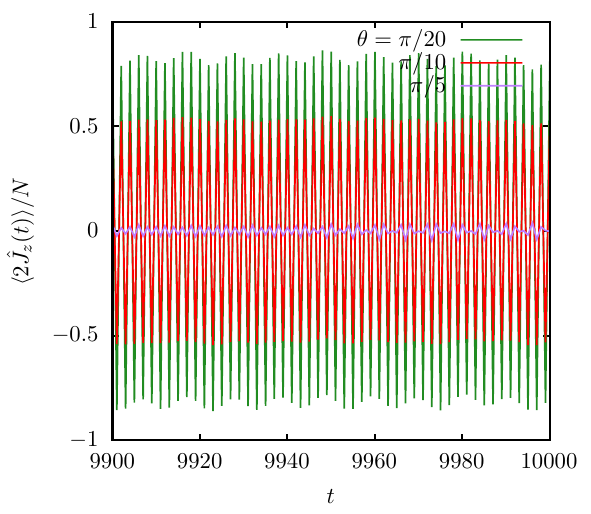}}\\
  (e) & (f) \\
 \includegraphics[width=68mm]{{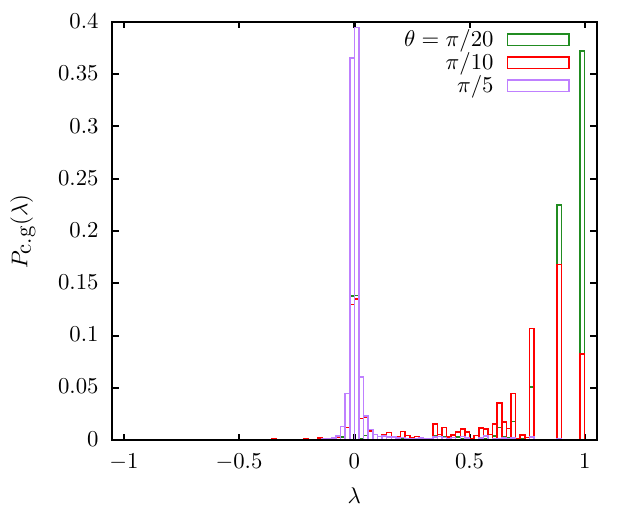}} &
 \includegraphics[width=68mm]{{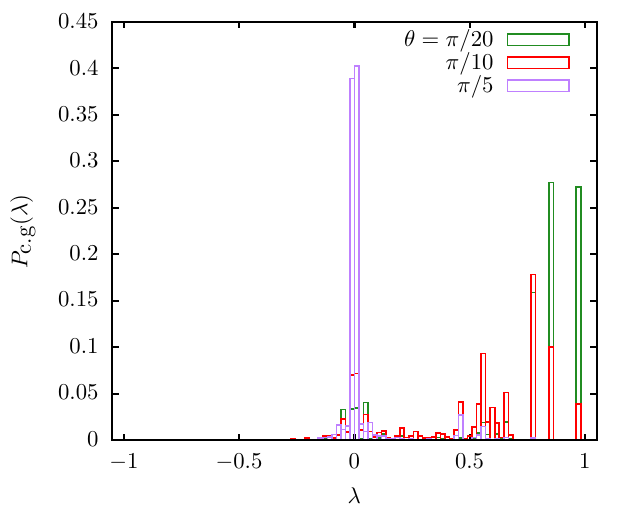}}
\end{tabular}
\caption{(a)-(b) Scatter plots where each yellow point corresponds to a Floquet-state doublet with reduced magnetization eigenvalues on the vertical axis and the logarithm of the $\pi$-shifted gap on the horizontal one. On the same plots we show $P_{\rm c.g}(\log_{10}\Delta^\pi$ [coarse-grained version of Eq.~\eqref{Pd:eqn}] for the three chosen initial states $\ket{\psi_{t}(\theta)}$ for $\theta=\pi/20$ (green), $\pi/10$ (red) and $\pi/5$ (purple) versus $\log_{10}\Delta_p^{\pi}$. (c)-(d) Time evolution of the $z$ magnetization with the two chosen initial states. The states with $\theta=\pi/20,\pi/10$ provide period-doubling oscillations. (e)-(f) $P_{\rm c.g.}(\lambda)$ [coarse-gained version of Eq.~\eqref{Pl:eqn}].  Numerical parameters for plots in the first column: $J=1,\,h_x=0.1,\,h_z=0.2,\,\alpha=2.2,\,\epsilon=0,\,N=19$. Numerical parameters for plots in the second column: $J=1,\,h_x=0.25,\,h_z=0.0,\,\alpha=0.6,\,\epsilon=0.1,\,N=19$.}\label{figZ:figtilted}
\end{center}
\end{figure*}
In order to perform a finite-size scaling, we take as initial state a so-called \textit{tilted} quantum state~\cite{PhysRevLett.117.090402}
\begin{equation}
    \ket{\psi_{\rm tilted}(\theta)}=\bigotimes_{i=1}^{N}[\cos(\theta)\ket{\uparrow_{i}}+\sin(\theta)\ket{\downarrow_{i}}].
\end{equation}
For our numerical simulations, we focus on the cases $\theta=\pi/5,\pi/10,\pi/20$. This initial state has the advantage that it is scalable with $N$, allowing for an appropriate finite-size scaling analysis. These initial states cannot be described in terms of the domain-wall theory in \cite{Liu2019}. This shows, as we can see below, that the time crystal behavior depends exclusively on the presence of ferromagnetic scars and on the overlap of these scarred states with the initial state, regardless of whether this initial state have magnetic domains or not. We remark that the expectation of the intensive magnetization on the tilted state is $\braket{\psi_{\rm titled}(\theta)|\hat{J}_z|\psi_{\rm tilted}(\theta)}/N = \cos(2\theta)$. This means that, when $\theta\ll 1$, the state has mainly overlap with magnetization eigenstates with a large eigenvalue, and then has large overlaps also with time translation symmetry breaking Floquet doublets, characterized by a large $\lambda$. So we expect to see time translation symmetry breaking physics for $\theta \ll 1$, as we are going to numerically check below.

Fig.~\ref{figZ:figtilted} is the analog of Fig.~\ref{figZ:fig} for the chosen tilted initial states. We can see that the $\log_{10}\Delta^\pi$ coarse-grained overlap distributions are broader, but there is anyway a clear difference between $\theta = \pi/10,\pi/20$ on one side and $\theta=\pi/5$ on the other. In the former case there are marked peaks at very small values of $\log_{10}\Delta^\pi$ and this corresponds to persistent period doubling oscillations [see Fig.~\ref{figZ:figtilted}(c,d)]. As before, the amplitude of the period doubling oscillations coincides with the mode of the $P_{\rm c.g.}(\lambda)$ distribution [see Fig.~\ref{figZ:figtilted}(e,f)], confirming the strict relation between period doubling and long-range order of the Floquet states with large overlap with the initial state. In $\pi/5$ case the amplitude of the period-doubling oscillations is very small, and correspondingly the peak of the $\lambda$ distribution is near 0.

\begin{figure*}
\begin{center}
  \begin{tabular}{cc}
  (a) & (b) \\
  \includegraphics[width=70mm]{{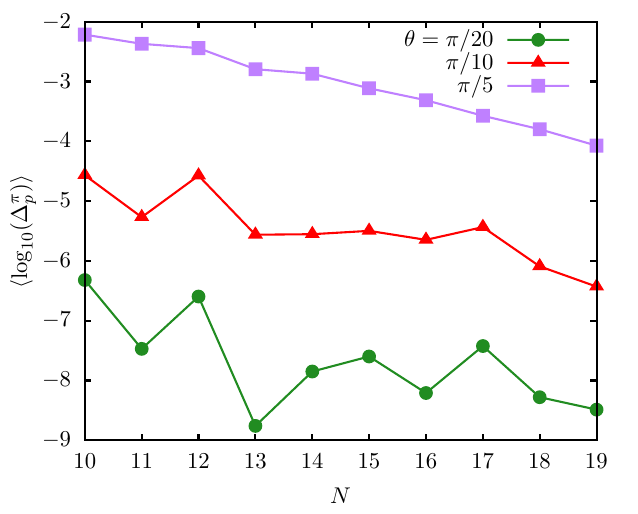}} &
  \includegraphics[width=70mm]{{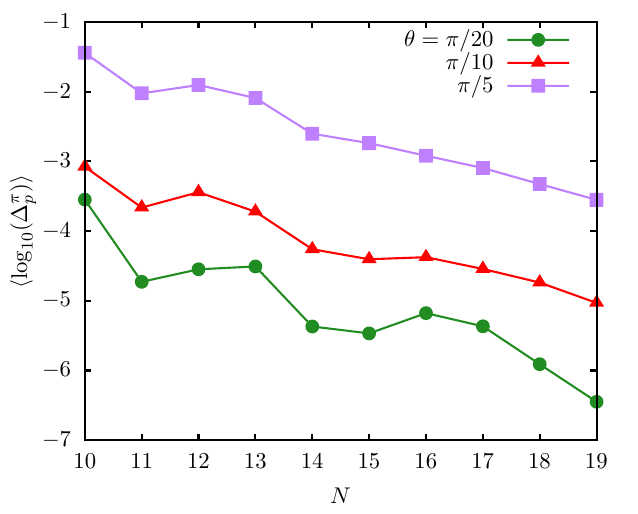}}\\
\end{tabular}
\caption{ $\braket{\log_{10}\Delta^\pi}$ [see Eq.~\eqref{lode:eqn}] versus $N$ for the two cases considered in Fig.~\ref{figZ:figtilted}. Model parameters are (a) $J=1,\,h_x=0.1,\,h_z=0.2,\,\alpha=2.2,\,\epsilon=0$, (b) $J=1,\,h_x=0.25,\,h_z=0.0,\,\alpha=0.6,\,\epsilon=0.1$. }\label{figZ:figscalingdelta}
\end{center}
\end{figure*}

In order to compare the behavior of different system sizes, we consider the average $\braket{\log_{10}\Delta^\pi}$ over the distribution of the $\pi$-shifted gap [see Eq.~\eqref{lode:eqn}]
and plot it versus $N$ for the cases considered in Fig.~\ref{figZ:figscalingdelta}. We see that it clearly decreases with the system size in a way that is not unreasonable to fit with a straight line. This suggests that increasing the system size, smaller $\pi$-shifted gaps become relevant for the dynamics, and the second term in Eq.~\eqref{Mnt:eqn} provides period-doubling oscillations persisting for a longer and longer time before dephasing occurs, a time whose order of magnitude is given by $t^*\sim 1/\exp\braket{\log_{10}\Delta^\pi}$, and is therefore exponential in the system size, as in other known time-crystal cases~\cite{PhysRevLett.117.090402,PhysRevB.109.174310,Surace_2019,PhysRevB.95.214307}.

\begin{figure*}
\begin{center}
  \begin{tabular}{cc}
  (a) & (b) \\
  \includegraphics[width=71mm]{{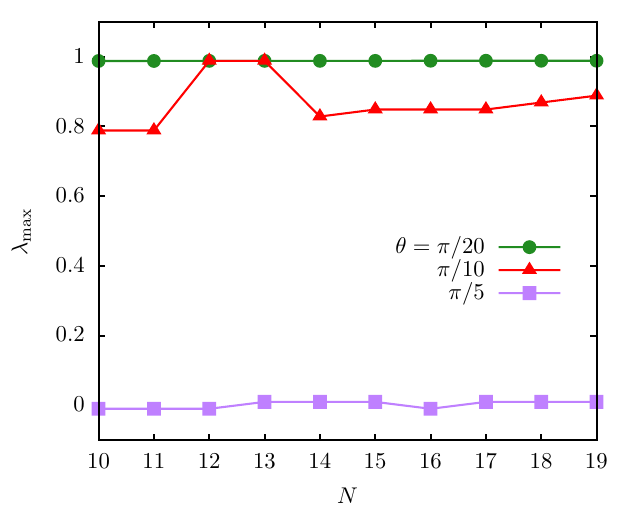}} &
  \includegraphics[width=71mm]{{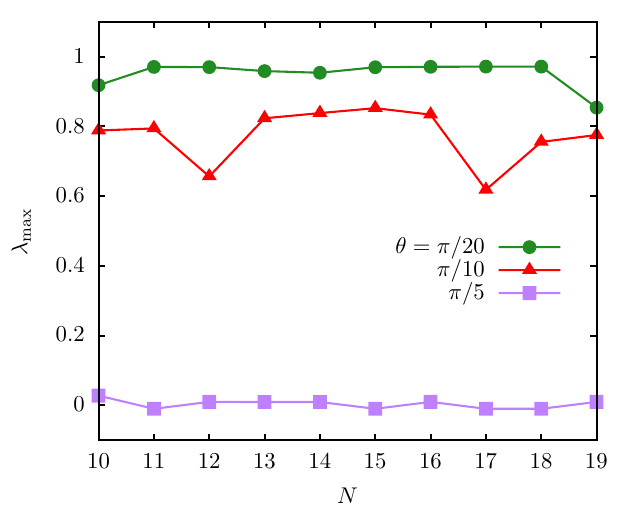}}\\
\end{tabular}
\caption{Mode of the coarse-grained normalized-eigenvalue distribution for the three chosen tilted states ($\theta=\pi/20,\pi/10,\pi/5$) and two different parameter choices. Model parameters are (a) $J=1,\,h_x=0.1,\,h_z=0.2,\,\alpha=2.2,\,\epsilon=0$, (b) $J=1,\,h_x=0.25,\,h_z=0.0,\,\alpha=0.6,\,\epsilon=0.1$. }\label{figZ:figscalinglambdamax}
\end{center}
\end{figure*}

We also show $\lambda_{\rm max}$ -- the mode of the coarse-grained $\lambda$ distribution -- versus $N$ for the considered cases in Fig.~\ref{figZ:figscalinglambdamax}. We see that for $\theta=\pi/20$ and $\theta=\pi/10$ it wiggles around a value significantly different from 0, suggesting that the same occurs for the amplitude of the period-doubling oscillations, which are indeed clearly visible as we have checked. The same does not occur for the larger $\theta=\pi/5$ where $\lambda_{\rm max}$ lies always in the surroundings of 0. Here the period-doubling oscillations are almost invisible.

\section{Conclusions} \label{sec:conclusions}
In conclusion we have studied a long-range Ising model periodically kicked with a spin flip perturbed so that it does not commute with the static part of the Hamiltonian. We have connected the presence of period-doubling oscillations to the properties of the Floquet states with large overlap with the initial state. To do that we have studied the time-translation symmetry breaking properties of the Floquet states. On one side we have studied the $\pi$-shifted gap, that probes how well a Floquet state has a partner with whom performs Rabi oscillations with period twice that of the driving. On the other side we have studied the normalized eigenvalues of the magnetization in each of these doublets, that probes how well the Floquet states are cat-state superpositions of long-range ordered states, and so how much the Rabi oscillations mentioned before are visible.

Choosing different initial states, some with domain walls, some with tilted spins, we have observed that persistent period doubling appeared when the Floquet states with relevant overlap with the initial state had small $\pi$-shifted gap and large normalized magnetization eigenvalues. We have performed this analysis looking at $\braket{\log_{10}\Delta^\pi}$ -- the average of the logarihtmic $\pi$-shifted gap overlap-weighted distribution -- and at $\lambda_{\rm max}$ -- the mode of the coarse-grained normalized-eigenvalue overlap-weighted distribution. The absolute value of $\lambda_{\rm max}$  corresponds to the amplitude of the period-doubling oscillations and $t^*=\exp\left(-\braket{\log_{10}\Delta^\pi}\right)$ provides the order of magnitude of the lifetime of these oscillations.

For the tilted states -- where a finite-size scaling is possible -- we observe that when there is period doubling $\lambda_{\rm max}$ is approximately constant with the system size and $\braket{\log_{10}\Delta^\pi}$ decreases in a way well described by a straight line, suggesting that the period-doubling oscillations persist for a lifetime exponential in the system size.

We remark that this persistent period doubling occurs even if there is time-translation symmetry breaking just for a part of the Floquet states. Indeed we find that only a minority of the Floquet states show time-translation symmetry breaking (being organized in doublets with small $\pi$-spectral pairing and displaying long range order). This can be seen first of all in the behavior of the average level spacing ratio whose value is always far from the Poisson one corresponding to a fully integrable dynamics. It can also be seen looking at the scatter plot of the entanglement entropy of the Floquet states versus the normalized magnetization eigenstates of the corresponding doublet.
The majority of the Floquet states has small magnetization eigenvalues and a large entanglement entropy, near to the Page value corresponding to random states and infinite-temperature thermal ones. This is perfectly consistent with a driven-system ETH picture~\cite{PhysRevE.90.012110,PhysRevB.101.064302}. Nevertheless a minority of eigenstates has a large normalized magnetization eigenvalue -- therefore long-range order -- and a small value of the entanglement entropy -- that's to say these states are locally non thermal. This situation is strictly similar to the one of quantum scars where a minority of nonthermal eigenstates of the Hamiltonian leads to nontrivial behavior in time for a certain class of initial states. Here it happens the same and the nontrivial behavior is the persistent period doubling. Quantum scarring in Floquet states has also been observed in~\cite{huang2025floquetquantummanybodyscars,PhysRevB.106.104302,PhysRevLett.129.133001,Maskara_2021,PhysRevB.102.224309,PhysRevB.108.205129}.

Although the nonthermal Floquet states are a minority, we nevertheless observe that their number is exponential in the system size and it is therefore not strange that we observe period doubling for many initial states. Similarly to the nondriven case of the same Hamiltonian (where the nontrivial behavior in time is a persistent nonvanishing magnetization rather than a period doubling) this exponential number of nonthermal eigenstates can be observed for values of the power-law exponent $\alpha$ well beyond the range corresponding to a mean-field behavior in the thermodynamic limit.

Future research perspectives include the application of our analysis to other cases of disorderless period doubling~\cite{PhysRevX.12.031037,PhysRevB.106.134301,Cae3851,Huang_2018,7f2b02b1-248f-4c3d-9457-7f1b889fec0e} where it is likely that just a part of the Floquet states shows time-translation symmetry breaking, with a quantum-scar picture similar to the one we have discussed here.

\acknowledgments
A. L. Corps and A. Relaño acknowledge financial support from acknowledges financial support by the Spanish grant PID2022-136285NB-C31 funded by Ministerio de Ciencia e Innovacion / Agencia Estatal de Investigación
MCIN/AEI/10.13039/501100011033 and FEDER “A Way of
Making Europe”. A. L. Corps also acknowledges support from the Czech Science Foundation under project No. 25-16056S and the JUNIOR UK Fund
project carried out at the Faculty of Mathematics and Physics,
Charles University. A. Russomanno acknowledges financial support from PNRR MUR Project PE0000023-NQSTI.

\begin{appendix}
\numberwithin{equation}{section}

\section{Symmetries of the Hamiltonian}\label{app:symmetries}
A convenient way to systematically incorporate all symmetries of the model is to view Eq. \eqref{eq:hamiltonian} with periodic boundary conditions as a closed ring of $N$ spins. In this geometry, the Hamiltonian is symmetric under the full set of transformations of the dihedral group $D_N$. This group is generated by a discrete rotation by $2\pi/N$, denoted by $\hat{\mathcal{R}}$, and by spatial inversion with respect to a symmetry axis of the regular $N$-gon, denoted by $\hat{\mathcal{I}}$. We construct our computational basis by explicitly exploiting these two symmetries, as detailed below.

\begin{enumerate}
\item \textit{Inversion symmetry.}  
We focus on the inversion operator $\hat{\mathcal{I}}$, choosing the symmetry axis that passes through the center of the ring. In the standard site basis,
$\ket{\phi}=\bigotimes_{i=1}^{N}\ket{\phi_i}_z$, with $\phi_i\in\{\uparrow,\downarrow\}$, this transformation corresponds to a reflection about the midpoint of the chain,
\begin{equation}
\hat{\mathcal{I}}\ket{\phi_1\,\phi_2\,\ldots\,\phi_N}_z
=
\ket{\phi_N\,\ldots\,\phi_2\,\phi_1}_z .
\end{equation}
We restrict our computational basis to the even-inversion subspace, consisting of states $\ket{\phi}$ that satisfy
$\hat{\mathcal{I}}\ket{\phi}=\ket{\phi}$.

\item \textit{Discrete rotational symmetry.}  
The generator of lattice translations is the rotation operator $\hat{\mathcal{R}}$, which acts on the site basis as
\begin{equation}
\hat{\mathcal{R}}\ket{\phi_1\,\phi_2\,\cdots\,\phi_N}_z
=
\ket{\phi_N\,\phi_1\,\cdots\,\phi_{N-1}}_z .
\end{equation}
Although $\hat{\mathcal{R}}$ and $\hat{\mathcal{I}}$ do not commute in general, they do commute within the zero- and $\pi$-momentum sectors. We exploit this property by constructing basis states with zero momentum and positive inversion parity, i.e.,
\[
\hat{\mathcal{R}}\ket{\phi}=\ket{\phi},
\qquad
\hat{\mathcal{I}}\ket{\phi}=\ket{\phi}.
\]
Other symmetry sectors are computationally more demanding and lead to the same qualitative behavior, so we do not consider them here.

\item \textit{Parity symmetry.}  
Finally, when $h_{z}=0$, the Hamiltonian in Eq.~\eqref{eq:hamiltonian} also conserves the global parity operator
\begin{equation}\label{eq:parity}
\hat{\Pi}=\prod_{i=1}^{N}\hat{\sigma}_i^x ,
\end{equation}
which commutes with both $\hat{\mathcal{I}}$ and $\hat{\mathcal{R}}$. Our computational basis therefore includes states from both parity sectors only when $h_{z}=0$. 
\end{enumerate}

\end{appendix}

\bibliography{biblio.bib}

\end{document}